\documentclass[reprint,aps,amsmath,amssymb,twoside,prd,showkeys,superscriptaddress]{revtex4-1}
\pdfoutput=1
\usepackage[colorlinks,citecolor=blue,urlcolor=blue,linkcolor=blue]{
hyperref}
\usepackage{graphicx}

\usepackage{latexsym}
\usepackage{amsfonts}
\usepackage{amssymb,amsmath}
\usepackage[english]{babel}

\usepackage{bm}
\usepackage{slashed}
\usepackage{dsfont}
\usepackage{braket}

\def\doi{http://doi.org}

\newcommand{\HCd}{\mathcal{H}}

\def\HCdt0{\tilde{\HCd}_{0}}

\def\bn{\mathbf{n}}
\def\bk{\mathbf{k}}

\def\mH{\mathcal{H}}
\def\mD{\mathcal{D}}
\def\mG{\mathcal{G}}
\def\mC{\mathcal{C}}
\newcommand{\tmD}{\tilde{\mathcal{D}}}

\newcommand{\affhel}{Department of Physics, University of Helsinki, P.O.
Box 64, FI-00014 Helsinki, Finland}
\newcommand{\affhip}{Helsinki Institute of Physics, P.O. Box 64,
FI-00014 University of Helsinki, Finland}
\newcommand{\affcam}{DAMTP, Centre for Mathematical Sciences,
University of Cambridge, Wilberforce Road, Cambridge CB3 0WA, UK}
\newcommand{\affcamPhys}{Institute of Astronomy, University of
Cambridge, Madingley Road, Cambridge, CB3 0HA, UK}

\begin{document}
\title{Mimetic Tensor-Vector-Scalar Cosmology: \\
Incorporating Dark Matter, Dark Energy and Stiff Matter}
\author{David Benisty}
\email{db888@cam.ac.uk}
\affiliation{\affcam}\affiliation{\affcamPhys}
\author{Moshe M. Chaichian}
\email{masud.chaichian@helsinki.fi}
\affiliation{\affhel}\affiliation{\affhip}
\author{Markku Oksanen}
\email{markku.oksanen@helsinki.fi}
\affiliation{\affhel}\affiliation{\affhip}

\begin{abstract}
Phenomenological implications of the Mimetic Tensor-Vector-Scalar theory
(MiTeVeS) are studied. The theory is an extension of the vector field
model of mimetic dark matter, where a scalar field is also incorporated,
and it is known to be free from ghost instability. In the absence of
interactions between the scalar field and the vector field, the obtained
cosmological solution corresponds to the General theory of Relativity
(GR) with a minimally-coupled scalar field. However, including an
interaction term between the scalar field and the vector field yields
interesting dynamics. There is a shift symmetry for the scalar field
with a flat potential, and the conserved Noether current, which is
associated with the symmetry, behaves as a dark matter component.
Consequently, the solution contains a cosmological constant, dark matter
and a stiff matter fluid. Breaking the shift symmetry with a non-flat
potential gives a natural interaction between dark energy and dark
matter.
\end{abstract}

\maketitle


\section{Introduction}
\label{sec:intro}
The Universe is found from many experiments and
observations to include 25\% of matter and 68\% of dark energy \cite{Riess:1998cb,Perlmutter:1998np,Abbott:2017wau,Aghanim:2018eyx}. There have been many models for dark matter \cite{Bertone:2004pz,Roszkowski:2017nbc,ArkaniHamed:2008qn,CyrRacine:2012fz,Foot:2014uba}. The description of dark energy, however, suffers from theoretical problems
\cite{Weinberg:1988cp,Sahni:2002kh,Copeland:2006wr,Cai:2009zp}), what gives motivation for modified theories of gravity \cite{Tsujikawa:2010zza,Capozziello:2011et,Nojiri:2017ncd}, as the General Relativity (GR) could not describe the data through the whole range of scales. One interesting class of modified theories of gravity are theories, where
a scalar field mimics the dark matter \cite{Chamseddine:2013kea} using a
Lagrange multiplier \cite{Golovnev:2013jxa,Chamseddine:2014vna} (see also \cite{Sebastiani:2016ras} for a review and \cite{Arroja:2015wpa,Arroja:2015yvd,Ganz:2018vzg,Casalino:2018tcd} for some developments of the mimetic theory). In its
original formulation, the mimetic theory of gravity can be obtained starting from the GR with a Lagrange multiplier that forces the kinetic part of the scalar field to behave as a constant in time~\cite{Barvinsky:2013mea}.
Consequently, the contribution of the mimetic field in the gravitational field equations is a pressureless perfect fluid. On a Friedmann-Robertson-Walker
(FRWM) metric this fluid behaves precisely as a dust component on the cosmological scales.
A minimal generalization of the mimetic dark matter was proposed by including a potential for the mimetic scalar field \cite{Chamseddine:2014vna}. The cosmological dynamics of the theory was further studied in \cite{Dutta:2017fjw}. The theory contains dynamic dark energy along with mimetic dark matter, and it was stressed that the early-time solution is not dominated by stiff matter unlike in simpler scalar-field models of dark energy, like in quintessence.

The Modified Newtonian Dynamics (MOND)
\cite{Milgrom:1983ca,Bekenstein:1984tv} is also an alternative successful
explanation for the flat rotation curves of galaxies. By violating the
Newton's second law at low accelerations, the Tully-Fisher relation is
recovered, without introducing an additional dark matter
~\cite{Tully:1977fu,McGaugh:2000sr,Chae:2020omu}. The Tensor-Vector-Scalar theory (TeVeS) is a covariant theory, which produces MOND in the
low energy limit \cite{Bekenstein:2004ne}. TeVeS is a stable theory and free from ghosts ~\cite{Chaichian:2014dfa}.
A new relativistic theory for MOND has been proposed
\cite{Skordis:2020eui} in order to reproduce the key cosmological observations, while still reproducing all the galactic and lensing phenomenology in the same way as TeVeS does.

%Heisenberg:2018acv,Heisenberg:2018wye,
Cosmology with vector fields has been considered widely in the
literature
\cite{Kase:2018nwt,Heisenberg:2018erb,Oliveros:2019zkl,
GallegoCadavid:2019zke,Kushwaha:2019rxj,Benisty:2018qed}. General
consideration of scalar-vector-tensor theories and their cosmological
solutions is given in \cite{Heisenberg:2018mxx}. A vector field model of
mimetic matter was proposed in \cite{Barvinsky:2013mea}. The consistency
of those theories with respect to the absence of ghost instability was studied in \cite{Chaichian:2014qba}, including a proposal of a new Tensor-Vector-Scalar gravity (MiTeVeS), which generalizes the previous models. The vector field was introduced to enable the possibility of rotating flows of mimetic matter and to avoid caustics of the geodesic flow \cite{Barvinsky:2013mea}. The scalar field in MiTeVeS, with its coupling to the vector field, has a significant impact on cosmology. In this paper we discover the homogeneous solution for this theory considering couple of possible interactions for the vector and scalar fields.

First in section~\ref{sec3} we confirm that the theory with the added
interactions does not suffer from pathologies like ghosts, gradient
instabilities or tachyons. The homogeneous solution is obtained in
sections~\ref{sec4} and \ref{sec5}. Cosmological constraints on the
solution are obtained in section~\ref{sec6}.
Interacting dark energy and dark matter are discussed in section~\ref{sec7}.

\section{The Theory}
\subsection{Lagrangian}
Gravitational theories with mimetic matter involve two metrics which are related by a conformal transformation.
The physical metric $\Phi^2g_{\mu\nu}$ is coupled minimally to all matter fields, where $\Phi$ is the conformal factor field that relates the physical metric to the metric $g_{\mu\nu}$. The Mimetic TeVeS theory includes a scalar field $\phi$ and a vector field $u_\mu$. The conformal factor is related to the scalar and vector fields by including a constraint with a Lagrange multiplier $\lambda$.
The Lagrangian is written as
\begin{equation}
\begin{split}
\label{S}
\mathcal{L} = \frac{1}{2} \Phi^2 \mathcal{R} + 3 \Phi^{,\alpha}
\Phi_{,\alpha} -\frac{1}{2} \lambda\left(\Phi^2+f(\phi)\, u^\mu u_\mu
\right)
\\
- \frac{\Phi^2}{2} \phi_{,\mu} \phi^{,\mu} - \Phi^4 V(\phi) -
\frac{\mu^2}{4} F_{\mu\nu}F^{\mu\nu} + \mathcal{L}_{\mathrm{m,int}},
\end{split}
\end{equation}
where $\mathcal{R}$ is the Ricci scalar determined by the metric $g_{\mu\nu}$, the kinetic term of the vector
field is defined in terms of
\begin{equation}
F_{\mu\nu} = \partial_\mu u_{\nu} - \partial_\nu u_{\mu},
\end{equation}
and the Lagrangian density
$\mathcal{L}_{\mathrm{m,int}}(\Phi^{2}g_{\mu\nu},\chi,\partial\chi,u_\mu
,\phi,\phi_{,\mu})$ represents matter fields
(denoted collectively as $\chi$) and all interactions among the
gravitational fields and the matter fields. Matter couples universally to the physical metric $\Phi^{2}g_{\mu\nu}$.

The concept of scale invariance is an attractive possibility for a
fundamental symmetry of nature. Dimensionless coupling constants for
example appear to be related to good renormalizability properties. In its most naive realizations, scale invariance is not a viable symmetry, since nature seems to have chosen some typical scales. The scale invariance can nevertheless be incorporated into realistic generally covariant field theories. Scale invariance has to be discussed in a more general framework than that of standard generally relativistic theories. In the MiTeVeS theory, the action incorporates the scale invariance
under the transformation
\begin{equation}
g_{\mu\nu} \rightarrow L^2 g_{\mu\nu} ,\quad  \Phi \rightarrow L^{-1}
\Phi,\quad \lambda \rightarrow L^{-2} \lambda,
\end{equation}
where $L = L(\vec{x},t)$ is an arbitrary function of space and time. The fields $\phi$ and $u_{\mu}$ are the invariant of the theory:
\begin{equation}
\phi \rightarrow \phi, \quad u_{\mu} \rightarrow u_{\mu}.
\end{equation}
This indicates that the theory is capable to have inflationary solutions naturally.

We can simplify the derivation of the field equations by fixing the
conformal symmetry of the theory. We shall choose the gauge condition as $\Phi=1$, which gives the theory in the Einstein frame form~\cite{Chaichian:2014qba}. This is equivalent to performing the following
conformal transformation in \eqref{S}:
\begin{equation}
g_{\mu\nu}\rightarrow\Phi^{-2}g_{\mu\nu}, \quad
\lambda\rightarrow\Phi^{2}\lambda.
\end{equation}
In this frame the Lagrangian is obtained without the auxiliary field
$\Phi$ as
\begin{equation}\label{action}
\begin{split}
\mathcal{L}= \frac{1}{2}\mathcal{R}(g_{\mu\nu})
-\frac{1}{2} \lambda\left( 1+f(\phi)\, u^\mu u_\mu \right)
- \frac{1}{2}\phi_{,\mu}\phi^{,\mu} \\
- V(\phi) -\frac{\mu^2}{4} F_{\mu\nu} F^{\mu\nu}
+\mathcal{L}_{\mathrm{m,int}},
\end{split}
\end{equation}
where $g_{\mu\nu}$ is now the physical metric that couples to matter
universally. This frame is equivalent for the setting
\begin{equation}
\Phi^2 = \frac{1}{8 \pi G} = 1
\end{equation}
where $G$ is the Newtonian Constant. We assume in the following analyses that the gravitational vector and scalar fields do not couple directly to matter fields:
\begin{equation}
\mathcal{L}_{\mathrm{m,int}}=
\mathcal{L}_{\mathrm{m}}(g_{\mu\nu},\chi,\partial\chi)
+\mathcal{L}_{\mathrm{int}}(g_{\mu\nu},u_\mu,\phi,\phi_{,\mu}).
\end{equation}
Furthermore, we consider the two simplest interactions for $\phi$ and $u_{\mu}$,
\begin{equation}\label{L_int.di}
\mathcal{L}_\mathrm{int}= - \frac{1}{2} h_1(\phi)u_{\mu}u^{\mu}
- \frac{1}{2} h_2(\phi) u^{\mu} \phi_{,\mu} ,
\end{equation}
which involve $u_\mu$ up to the second power and are at most linear in
$\phi_{,\mu}$.
Terms that involve higher powers of $\phi_{,\mu}$ would modify the
kinetic term of the scalar field $\phi$, which will not be considered in this work.
These assumptions on $\mathcal{L}_{\mathrm{m,int}}$ are made here mainly for simplicity.
More generally, interactions of higher order as well as couplings between matter and the scalar and/or vector fields are possible.
We also note that quantum corrections may induce such terms, when there is no symmetry to prevent it.
% Also notice that these terms are not coupled to the scalar $\lambda$.

\subsection{Field equations}
Varying the Lagrangian \eqref{action} minimally coupled to the
interaction term \eqref{L_int.di} with respect to $g^{\mu\nu}$,
$\lambda$, $\phi$ and $u_\mu$ gives the field equations as
\begin{subequations}
\begin{equation}
G_{\mu\nu}=\lambda f(\phi)u_\mu u_\nu +T^\phi_{\mu\nu}
+T^u_{\mu\nu}+T_{\mu\nu}^{\mathrm{int},1}+T_{\mu\nu}^{\mathrm{int},2}+T_
{\mu\nu},\label{Einstein}
\end{equation}
\begin{equation}
 f(\phi)\, u^\mu u_\mu=-1,\label{constraint}
\end{equation}
\begin{equation}
\Box \phi + \frac{1}{2} h_2(\phi) \nabla_{\mu}u^{\mu} = V'(\phi) +
\frac{1}{2}\left(\lambda f'(\phi) +
h_1'(\phi)\right)u^{\mu}u_{\mu},\label{phi.feq}
\end{equation}
\begin{equation}
\mu^2\nabla_\nu F^{\nu\mu}=\left[\lambda f(\phi)
+ h_1(\phi)\right] u^{\mu} + \frac{1}{2} h_2(\phi) \phi^{,\mu}
,\label{u.feq}
\end{equation}
\label{fieldequations}
\end{subequations}
where in the modified Einstein equation \eqref{Einstein} we have
defined the energy-momentum tensors as:
\begin{equation}
T^\phi_{\mu\nu}=\phi_{,\mu}\phi_{,\nu} -g_{\mu\nu}\left(
\frac{1}{2}\phi_{,\alpha}\phi^{,\alpha} +V(\phi) \right),
\end{equation}
\begin{equation}
T^u_{\mu\nu}=\mu^2\left( F_{\mu\rho}F_{\nu}^{\phantom\nu\rho}
-\frac{1}{4}g_{\mu\nu}F_{\rho\sigma}F^{\rho\sigma} \right),
\end{equation}
\begin{equation}\label{T_munu}
T_{\mu\nu}=-\frac{2}{\sqrt{-g}}\frac{\delta}{\delta g^{\mu\nu}}
\int
d^4x\sqrt{-g}\,\mathcal{L}_{\mathrm{m}}(g_{\mu\nu},\chi,\partial\chi),
\end{equation}
\begin{equation}
T_{\mu\nu}^{\mathrm{int},1} = h_1(\phi) \left( u_{\mu} u_{\nu} -
\frac{1}{2} g_{\mu\nu}\left(
u^{\alpha} u_{\alpha} \right) \right),\label{T^int1}
\end{equation}
\begin{equation}
T_{\mu\nu}^{\mathrm{int},2} = h_2(\phi) \left( u_{(\mu} \phi_{,\nu)} -
\frac{1}{2} g_{\mu\nu}\left(u^\alpha
\phi_{,\alpha}\right)\right),\label{T^int2}
\end{equation}
where the parentheses around indices denote symmetrization:
\begin{equation}
 u_{(\mu} \phi_{,\nu)} = \frac{1}{2}  \left( u_{\mu} \phi_{,\nu} +
u_{\nu} \phi_{,\mu}\right),
\end{equation}
which ensures the symmetric form of the energy momentum tensor.
The contributions of the interactions between the scalar field and the
vector field to the energy-momentum tensor are given in \eqref{T^int1}
and \eqref{T^int2}.

\section{Healthiness of the theory}\label{sec3}
In this section we discuss the absence of ghosts, propagation of
perturbations and stability, particularly regarding the interactions
that we have introduced in \eqref{L_int.di}. First we complete the
Hamiltonian analysis \cite{Chaichian:2014qba} with the interactions
\eqref{L_int.di}.
The structure of the Hamiltonian analysis remains largely unchanged,
since the interactions \eqref{L_int.di} involve only one derivative of
the scalar field. Hence we only present the significant changes to the
Hamiltonian analysis achieved in \cite{Chaichian:2014qba}.
We have performed the Hamiltonian analysis of MiTeVeS with the usual method suited for constrained systems \cite{Gitman:1990qh,Henneaux:1992ig,Chaichian:2001da}.

In the last subsection we discuss the relation of MiTeVeS and the Einstein-aether theory of gravity, particularly regarding the requirements of stability.

\subsection{Hamiltonian analysis}\label{sec3.1}
We assume foliation of spacetime to a union of spacelike hypersurfaces and the Arnowitt--Deser--Misner parameterization of the metric,
\[
ds^2=-N^2dt^2+h_{ij}(N^idt+dx^i)(N^jdt+dx^j).
\]
The variables of the Hamiltonian formulation are the lapse $N$, the shift $N^i$, the induced metric $h_{ij}$ on the spatial hypersurface, the scalar fields $\lambda$, $\Phi$ and $\phi$, and the components of the vector field $u_i$ and $u_{\bn}$ that are tangent and normal to the spatial hypersurfaces, respectively. The canonical momenta conjugate to those variables are $\pi_N$, $\pi_i$, $\pi^{ij}$, $p_\lambda$, $p_\Phi$, $p_\phi$, $p^i$ and $p_{\bn}$, respectively. The covariant derivative determined by the metric $h_{ij}$ is denoted with $D$, and $R$ is the corresponding Ricci scalar.

We obtain the Hamiltonian as
\begin{multline}
H=\int d^3x\,\Bigl( N\mH_T+N^i\mH_i+v_\mD \mD+v_N\pi_N
+v^i\pi_i \\
+v_\lambda p_\lambda+v_\bn p_\bn \Bigr) ,\label{H}
\end{multline}
where
\begin{align}
\mH_T&=\frac{2}{\sqrt{h}\Phi^2}\pi^{ij}\mG_{ijkl}\pi^{kl}-
\frac{1}{2}\sqrt{h}R\Phi^2 +\partial_i
[\sqrt{h}h^{ij}\partial_j\Phi^2] \nonumber \\
&-3\sqrt{h}h^{ij}\partial_i\Phi
\partial_j\Phi-\frac{1}{2}\sqrt{h}\lambda\bigl(\Phi^2
+f(\phi)u_i h^{ij}u_j \nonumber \\
&-f(\phi)u_\bn^2\bigr) +\frac{1}{2\mu^2\sqrt{h}} p^ih_{ij}p^j  \nonumber \\
&+\frac{\mu^2}{4}\sqrt{h}
h^{ik}h^{jl}(D_iu_j-D_ju_i)(D_ku_l-D_lu_k) \nonumber \\
&-u_\bn D_i p^i +\frac{1}{2}\sqrt{h}\Phi^{2}h_1(\phi)\left(
u_ih^{ij}u_j-u_{\bn}^{2} \right) \nonumber\\
&+\frac{1}{8}\sqrt{h}\Phi^2\left( h_2(\phi)u_{\bn} \right)^2
+\frac{p_\phi^2}{2\sqrt{h}\Phi^2}
-\frac{1}{2}h_2(\phi)u_{\bn}p_\phi  \nonumber\\
&+\frac{1}{2}\sqrt{h}\Phi^2 \partial_i\phi \,h^{ij} \left(
\partial_j\phi
+h_2(\phi)u_j \right)
+\sqrt{h}\Phi^4 V(\phi) ,\label{H_T} \\
\mH_i&=p_\Phi\partial_i\Phi-2h_{ik}D_j \pi^{kj} +\partial_i u_j
p^j-\partial_j (u_i p^j)+p_\phi\partial_i\phi ,\\
\mD&=p_\Phi\Phi-2\pi^{ij}h_{ij}.
\end{align}
In \eqref{H} all the fields denoted with $v$ are Lagrange multipliers.
We have denoted the De Witt metric in \eqref{H_T} as
\[
\mG_{ijkl}=\frac{1}{2}(h_{ik}h_{jl}+h_{il}h_{jk})-\frac{1}{2}h_{ij}h_{kl}.
\]
The primary constraints are
\begin{align}
p_\lambda&\approx 0 , \quad  \pi_N\approx 0 , \quad  \pi_i \approx
0 , \quad  p_{\bn}\approx 0, \nonumber\\
\tmD&=p_\Phi\Phi-2\pi^{ij}h_{ij}+2p_\lambda\lambda+p_{\bn}u_{\bn}\approx
0.
\end{align}
The preservation of the primary constraints $\pi_N$ and $\pi^i$ during
the
time evolution of the system implies following secondary constraints
\begin{equation}
\mH_T\approx 0 ,\quad \mH_i\approx 0.
\end{equation}
The requirement of the preservation of the constraint
$p_\lambda\approx 0$ implies a secondary constraint,
\begin{equation}\label{C_lambda}
\mC_\lambda=\sqrt{h}\left(\Phi^2+f(\phi)u_i h^{ij}u_j
-f(\phi)u_{\bn}^2\right) \approx 0.
\end{equation}
The preservation of the constraint $p_{\bn}$ gives a secondary
constraint,
\begin{multline}
\mC_\bn=\sqrt{h}\left( \Phi^2h_1(\phi) -\lambda f(\phi)
-\frac{1}{4}\Phi^2h_2^2(\phi) \right)u_{\bn}\\
+D_i p^i +\frac{1}{2}h_2(\phi)p_\phi\approx 0 .
\end{multline}
The constraint $\tmD$ is the first class constraint that generates the
scaling transformation. $\mH_T$, $\pi_N$, $\mH_i$ and $\pi_i$ are the
first class constraints which reflect of the diffeomorphism invariance
of the theory. The four remaining constraints $p_\lambda$, $\mC_{\bn}$,
$p_{\bn}$ and $\mC_{\lambda}$ are the second class constraints,
which can be interpreted as follows. The constraints $p_\lambda$ and $\mC_{\bn}$ can be used to fix the pair of canonical variables $\lambda$ and $p_\lambda$, while the constraints $p_{\bn}$ and $\mC_{\lambda}$ can be used to fix the variables $u_\bn$ and $p_{\bn}$.
Overall the Hamiltonian structure remains similar to \cite{Chaichian:2014qba}.

There is no sign of presence of ghosts in the Hamiltonian.

\subsection{Propagation of perturbations}
We consider the linearization of the theory. The tensor, vector and
scalar fields are written as
\begin{equation}
g_{\mu\nu}=g^{(0)}_{\mu\nu}+\bar{g}_{\mu\nu},\quad
u_{\mu}=u^{(0)}_{\mu}+\bar{u}_{\mu},\quad
\phi=\phi^{(0)}+\bar\phi,
\end{equation}
where $g^{(0)}_{\mu\nu}$, $u^{(0)}_{\mu}$ and $\phi^{(0)}$ are the
background and $\bar{g}_{\mu\nu}$, $\bar{u}_{\mu}$ and $\bar\phi$ are
small amplitude perturbations with a length scale of variations much
shorter than the length scale of variation of the background.
We study the propagation of perturbations

For simplicity we consider the Minkowski background:
$g^{(0)}_{\mu\nu}=\eta_{\mu\nu}$, $\phi^{(0)}=\phi_0$ (constant) and
$u^{(0)}_{\mu}=(f(\phi_0)^{-\frac{1}{2}},0,0,0)$. This background can be
obtained in the limit of vanishing mimetic energy-momentum in the
background: $V(\phi_0)=-\lambda/2$ and $h_1(\phi_0)=-\lambda f(\phi_0)$.

We consider the field equations \eqref{fieldequations} to lowest order
in the perturbations.
The Einstein field equations \eqref{Einstein} are linearized as
$\bar{G}_{\mu\nu}=T^{tot}_{\mu\nu}$, where $\bar{G}_{\mu\nu}$ is the
linearized Einstein tensor and $T^{tot}_{\mu\nu}$ is the total
energy-momentum, where the scalar and vector fields appear like
additional matter components. The propagation of the metric
perturbations is similar to GR.
The rest of the field equations give the linearized constraint
\begin{equation}
-\frac{f'(\phi_0)}{f(\phi_0)}\bar\phi+\bar{g}^{00}-2f(\phi_0)^{\frac{1}{
2}}\bar{u}_{0}=0,\label{linconstr}
\end{equation}
and the linearized field equations for the scalar and vector fields
\begin{subequations}
\begin{align}
\Box\bar\phi + \frac{1}{2} h_2 \partial_{\mu}\bar{u}^{\mu} &=
B\bar\phi,\label{phi.lin}\\
\mu^2\left( \Box\bar{u}^{\mu} -\partial^\mu\partial_\nu\bar{u}^{\nu}
\right) &=\frac{1}{2} h_2(\phi_0) \bar\phi^{,\mu},\label{u.lin}
\end{align}\label{phi+u.lin}
\end{subequations}
where we have defined
\begin{equation}
B=V''(\phi_0) -\left(\frac{\lambda f' + h_1'}{2f}\right)'(\phi_0).
\end{equation}
For the constant functions \eqref{constfunc} considered in
Sec.~\ref{sec5} we would have $B=0$.
The constraint \eqref{linconstr} determines one of the three involved
perturbations in terms of the other two. Differentiating the vector
equation \eqref{u.lin} we obtain $\Box\bar\phi=0$, which according to
the scalar equation \eqref{phi.lin} implies
$h_2(\phi_0)\partial_\mu\bar{u}^{\mu}=2B\bar\phi$; in the case of the
constant functions \eqref{constfunc} the perturbation of the vector
field is transverse.
Hence the scalar and vector equations \eqref{phi+u.lin} simplify as
\begin{subequations}
\begin{align}
\Box\bar\phi&=0,\label{phi.lin2}\\
\Box\bar{u}_{\mu}&=\frac{C}{\mu^2}\bar\phi_{,\mu},\label{u.lin2}
\end{align}\label{phi+u.lin2}
\end{subequations}
where we have defined
\begin{equation}
C=\frac{1}{2} h_2(\phi_0)+\frac{2\mu^2B}{h_2(\phi_0)}.
\end{equation}
The functions $V$, $f$, $h_1$ and $h_2$ are chosen so that $C\ge0$.

The perturbations of the tensor and scalar fields satisfy standard wave
equations and propagate with the speed of light ($c=1$).
The perturbation $\bar{u}_{\mu}$ of the vector field satisfies the
forced wave equation \eqref{u.lin2}, where the forcing term is the
gradient of the scalar perturbation. The solution of \eqref{u.lin2} is a
sum of the solution of the homogeneous wave equation
$\Box\bar{u}_{\mu}=0$ and the solution of the inhomogeneous wave
equation, $\bar{u}_{\mu}=\bar{u}_{\mu}^{H}+\bar{u}_{\mu}^{I}$. The
latter can be written with a Green's function as
\begin{equation}
\bar{u}_{\mu}^{I}=\frac{C}{\mu^2}\int d^4y\,G(x-y)\bar\phi_{,\mu}(y),
\end{equation}
where the Green's function $G$ satisfies $\Box G(x)=\delta(x)$ with
suitable boundary conditions.
Consider the plane wave solutions to the homogeneous wave equations
$\bar\phi=\epsilon e^{ik_{\mu}x^{\mu}}$ and
$\bar{u}_{\mu}^{H}=\epsilon_{\mu}e^{ik_{\mu}x^{\mu}}$. Then the solution
of \eqref{u.lin2} takes the form
\begin{equation}
\begin{split}
\bar{u}_{\mu}&=\epsilon_{\mu}e^{ik_{\mu}x^{\mu}}
+i\epsilon k_{\mu}\frac{C}{\mu^2}\int d^4y\,G(x-y)e^{ik_{\mu}y^{\mu}}\\
&=\left(
\epsilon_{\mu}+i\epsilon\frac{C}{\mu^2}\frac{k_{\mu}}{k^2+i\varepsilon}
\right) e^{ik_{\mu}x^{\mu}},\label{u.sol}
\end{split}
\end{equation}
where we have used that the Fourier transformation $\hat G(k)$ of the
Green's function satisfies $k^2\hat G(k)=-1$. This solution tells us
three significant things. The solution \eqref{u.sol} is a plane wave
with a $k$-dependent amplitude and it gives the dispersion relation
$k^2=0$ by substitution into \eqref{u.lin2}. The $k$-dependent part of
the amplitude of \eqref{u.sol} is parallel to $k_{\mu}=(\omega,\bk)$,
which means that if we write
$\bar{u}_\mu=(\bar{u}_0,\bm{\bar{u}}^{\perp}+\bm{\bar{u}}^{\parallel})$,
where $\bm{\bar{u}}^{\perp}$ is perpendicular to $\bk$ and
$\bm{\bar{u}}^{\parallel}$ is parallel to $\bk$, the perpendicular
component
$\bm{\bar{u}}^{\perp}=\bm{\epsilon}^{\perp}e^{ik_{\mu}x^{\mu}}$, i.e.,
only the components $\bar{u}_0$ and $\bm{\bar{u}}^{\parallel}$ have
$k$-dependent amplitudes. Thirdly, the scale of variation of the
amplitude is given by the parameter $C/\mu^2$.

The analysis of perturbations on a general curved background is somewhat
longer. The key point to observe about the field equations
\eqref{fieldequations} is that the kinetic matrix of the Lagrangian is
diagonal with respect to the tensor, vector and scalar fields, and
furthermore it is a direct sum of the Einstein, Klein-Gordon and Maxwell
kinetic terms, respectively. Therefore, as long as the functions $V$,
$f$, $h_1$ and $h_2$ are chosen reasonably (as demonstrated above),
there are no instabilities or tachyons.

Finally, we remark that the interaction terms we have introduced in
\eqref{L_int.di} represent attractive interactions between the vector
and scalar fields. In the linearized theory, the interaction terms
$-\frac{1}{2}h_1\bar{u}_{\mu}\bar{g}^{\mu\nu}\bar{u}_{\nu}
-\frac{1}{2}h_2\bar{u}_{\mu}\bar{g}^{\mu\nu}\bar\phi_{,\nu}$
describe an interaction of the vector field with itself and an
interaction between the vector and scalar fields. Both interactions are
mediated by the tensor field, i.e., the graviton, which is a massless
spin-2 field. It is well known that an interaction mediated by an even
integer spin field is always attractive. Such an attractive interaction
cannot drive the system to an illimitably excited state. So it is clear
that the addition of these interactions could not have introduced
instabilities.

\subsection{Relation to Einstein-aether theory and potential issues with stability}
When the function $f(\phi)$ is a constant, the vector field $u_\mu$ is similar to the aether vector field of Einstein-aether theory \cite{Jacobson:2000xp}.
In Einstein-aether theory the vector field is constrained to be a timelike unit vector, $u_\mu u^\mu=-1$.
Einstein-aether theory has been studied extensively as a way to consider the consequences of dynamical breaking of local Lorentz invariance such as variation of the speed of light or high frequency dispersion.
In our theory, the vector field is coupled to the scalar field $\phi$ in the Lagrangian \eqref{action}. In this work we only consider the two interactions in \eqref{L_int.di} in addition to the constaint \eqref{constraint}.

The stability of the aether vector field on Minkowski background spacetime  was studied in \cite{Carroll:2009em}.
It was found that only three kinetic terms have no linear instabilities or negative-energy ghosts: the sigma model $\partial_\mu u_\nu\partial^\mu u^\nu$, the Maxwell term $F_{\mu\nu}F^{\mu\nu}$, and the scalar kinetic term $(\partial_\mu u^\mu)^2$.
Only for the the sigma model the Hamiltonian of the vector field on the Minkowski background is bounded from below.

The Hamiltonian density of a field on a fixed background should not be confused with the Hamiltonian density of the full gravitational theory. The latter is a sum of constraints, as can be seen in the present case in \eqref{H}, and thus it is always zero.
The total energy of the system can be determined in terms of the boundary terms of the Hamiltonian (similarly as for General Relativity \cite{Hawking:1995fd}).
Positivity of the total energy for Einstein-aether theory was shown in \cite{Garfinkle:2011iw} for a class of solutions where the vector field is orthogonal to the spatial hypersurfaces of constant time and has a vanishing divergence on such hypersurfaces. We do not consider the boundedness of the total energy in this work.

Instead, similar to the analysis of \cite{Carroll:2009em}, we consider boundedness of the Hamiltonian of the vector field and the scalar field on the Minkowski background. In other words we fix the metric and ignore its dynamics. Then the Hamiltonian $H=\int d^3x\mH$ is obtained as
\begin{equation}
\begin{split}
\mH&=\frac{1}{2\mu^2} p^ip^i +\frac{1}{2}p_\phi^2
-u_\bn \partial_i p^i -\frac{1}{2}h_2(\phi)u_{\bn}p_\phi \\
&+\frac{\mu^2}{4}(\partial_iu_j-\partial_ju_i)(\partial_iu_j-\partial_ju_i) \\
&+\frac{1}{8}\left( h_2(\phi)u_{\bn} \right)^2
+\frac{1}{2}\partial_i\phi \left( \partial_i\phi
+h_2(\phi)u_i \right) +\tilde V(\phi).
\end{split}
\end{equation}
where we have also gauge fixed the conformal symmetry with $\Phi=1$, used the second class constraints, and redefinied the potential of the scalar field as
$\tilde V(\phi)=V(\phi)+\frac{h_1(\phi)}{2f(\phi)}$.
Completing the squares in the Hasimilarly miltonian density gives
\begin{equation}\label{H.vectorscalar}
\begin{split}
\mH&=\frac{1}{2\mu^2} \left( p^i+\mu^2\partial_iu_\bn \right)^2
-\frac{\mu^2}{2}\left( \partial_iu_{\bn} \right)^2 \\
&+\frac{\mu^2}{4}\left(\partial_iu_j-\partial_ju_i\right)^2
+\frac{1}{2}\left( p_\phi -\frac{1}{2}h_2(\phi)u_{\bn} \right)^2 \\
&+\frac{1}{2}\left( \partial_i\phi +\frac{1}{2}h_2(\phi)u_i \right)^2
-\frac{1}{8}\left( h_2(\phi)u_i \right)^2 \\
&+\tilde V(\phi).
\end{split}
\end{equation}
Note that $u_\bn$ is fixed by the constraint \eqref{C_lambda} as
\begin{equation}\label{constraint.Minkowski}
u_\bn^2=\frac{1}{f(\phi)}+u_iu_i
\end{equation}
and the definitions of the canonical momenta are now
\begin{equation}
p^i=\mu^2\left( \partial_tu_i-\partial_iu_{\bn} \right),\quad
p_\phi=\partial_t\phi.
\end{equation}
There are two negative terms in the Hamiltonian \eqref{H.vectorscalar}, namely, $-\frac{\mu^2}{2}\left( \partial_iu_{\bn} \right)^2$ and $-\frac{1}{8}\left( h_2(\phi)u_i \right)^2$.
The former term comes from the Maxwell kinetic term, while the latter is due to the second interaction term between $\phi$ and $u_\mu$ in  \eqref{L_int.di}.
In the previous subsection we showed that the linear perturbations of the scalar and vector fields do not grow without a limit on this background.
Perturbations on other backgrounds might exhibit stability problems.
For instance, if there exist a solution where $h_2^2(\phi)u_iu^i$ grows to become arbitrarily large, the Hamiltonian of the vector and scalar fields is not bounded from below.
Thus, solutions with instability might exist.

Stability of spherically symmetric solutions in Einstein-aether theory was studied in \cite{Seifert:2007fr}. It was found that the spherically symmetric perturbations are stable only if the coupling constant of the Maxwell kinetic term satisfies $\mu^2>4$, i.e. the coupling in the Lagrangian satisfies $-\frac{\mu^2}{4}<-1$. Furthermore, in order to avoid noticeable changes in tidal effects it is also necessary to include a further kinetic term or terms in addition to the Maxwell term. The coupling constants of the kinetic terms are severely constrained by the tidal effects. In particular, we could include the scalar kinetic term $(\nabla_\mu u^\mu)^2$ with the value of its coupling constant set either very close to $1$ or very close to $\frac{\mu^2}{4}$. Since MiTeVeS has the similar vector field (with the same kinetic structure and a similar contraint), we have confirmed that the coupling constants of the kinetic terms must be constrained similarly as in Einstein-aether gravity.

Spatially anisotropic cosmology in Einstein-aether theory, where the vector field has a nonvanishing spatial vacuum expectation value, was studied in \cite{Himmetoglu:2008zp}. A longitudinal polarization of the linear perturbation of the vector field was found to become a ghost at the horizon crossing. Anisotropic cosmology has not been considered in MiTeVeS and it is beyond the scope of the present work.

\section{Homogeneous Solution}\label{sec4}
In order to find the behavior of the theory for our universe, we consider a flat FLRW metric:
\begin{equation}
\mathrm{d}s^2=-dt^2+a(t)^2 (dx^2 + dy^2 + dz^2),
\end{equation}
and a time-dependent homogeneous ansatz for the scalar field
\begin{equation}
\phi=\phi(t).
\end{equation}
For the vector field we also consider a time-dependent homogeneous ansatz:
\begin{equation}\label{u^(0)}
u_{\mu}=(A(t),0,0,0).
\end{equation}
The energy-momentum tensor \eqref{T_munu} contains the contribution of a matter fluid or fluids as:
\begin{equation}
T^{m\mu}_{\phantom{m\mu}\nu}
=\mathrm{diag}(-\rho_m,p_m,p_m,p_m).
\end{equation}
The constraint \eqref{constraint} reduces to
\begin{equation}\label{constraint:Cosmo}
f(\phi)A^2=1.
\end{equation}
Since we assume a homogeneous solution, the kinetic term of the vector field that contains an anti-symmetric part, vanishes $F_{\mu\nu} = 0$.
Eq. (\ref{u.feq}) with the cosmological background gives:
\begin{equation}\label{vector:Cosmo}
\lambda f(\phi)A+h_1(\phi)A +\frac{1}{2}h_2(\phi)\dot{\phi}=0,
\end{equation}
The variation with respect the scalar field $\phi$, Eq. (\ref{phi.feq}), with a cosmological background, yields:
\begin{equation}\label{scalar:Cosmo}
\begin{split}
-\frac{1}{2} A^2 \left(\lambda f'(\phi)+h_1'(\phi)\right)
+\frac{1}{2}\dot{A} h_2(\phi)\\+\frac{3}{2} HA h_2(\phi)
+V'(\phi)+3H\dot{\phi}+\ddot{\phi}=0.
\end{split}
\end{equation}
The Einstein tensor with the energy-momentum tensor terms \eqref{Einstein} gives:
\begin{subequations}
\begin{equation}\label{Friedmann1}
3H^2=\lambda +\frac{1}{2}A^2 h_1(\phi)+\frac{1}{2}A h_2(\phi)
\dot{\phi}+V(\phi)+\frac{1}{2} \dot{\phi}^2 + \rho_m
\end{equation}
and
\begin{equation}\label{Friedmann2}
-3H^2 -2\dot{H}=\frac{1}{2}A^2 h_1(\phi)+\frac{1}{2}A h_2(\phi)
\dot{\phi}-V(\phi)+\frac{1}{2} \dot{\phi}^2 + p_m.
\end{equation}
\end{subequations}
We solve $A$ and $\lambda$ in terms of $\phi$ and $\dot\phi$ from
\eqref{constraint:Cosmo} and \eqref{vector:Cosmo} as
\begin{equation}\label{A.and.lambda}
\begin{split}
 A=\frac{1}{\sqrt{f(\phi)}},\quad  \lambda=-\frac{h_1(\phi)}{f(\phi)}-\frac{h_2(\phi)\dot\phi}{2\sqrt{f(\phi)}},
\end{split}
\end{equation}
where we assume that $A$ and $f(\phi)$ are positive, and obtain
$\dot{A}$ by differentiating \eqref{constraint:Cosmo}:
\begin{equation}\label{dotA}
f'(\phi)\dot{\phi}A^2+2f(\phi)A\dot{A}=0\quad\Rightarrow\quad
\dot{A}=-\frac{f'(\phi)\dot{\phi}}{2f(\phi)^{\frac{3}{2}}}.
\end{equation}
Then we insert \eqref{A.and.lambda} and \eqref{dotA} into
\eqref{scalar:Cosmo}, \eqref{Friedmann1} and \eqref{Friedmann2},
which gives:
\begin{subequations}
\begin{equation}\label{scalar:Cosmo.2}
\begin{split}
\frac{f'(\phi)h_1(\phi)}{2f(\phi)^2} -\frac{h'_1(\phi)}{2f(\phi)}
+V'(\phi)\\+\frac{3Hh_2(\phi)}{2\sqrt{f(\phi)}}
+3H\dot{\phi}+\ddot{\phi}=0,
\end{split}
\end{equation}
\begin{equation}\label{Friedmann1.2}
3H^2=-\frac{h_1(\phi)}{2f(\phi)}+V(\phi)+\frac{1}{2} \dot{\phi}^2
+\rho_m,
\end{equation}
\begin{equation}\label{Friedmann2.2}
-3H^2-2\dot{H}=\frac{h_1(\phi)}{2f(\phi)}-V(\phi)
+\frac{h_2(\phi)\dot{\phi}}{2\sqrt{f(\phi)}}
+\frac{1}{2} \dot{\phi}^2 + p_m.
\end{equation}
\end{subequations}
These equations summarize the homogeneous evolution of the universe considering only the scalar field $\phi$ as an external degree of freedom besides the scale parameter of the universe. One can check the consistency of the equations using the covariant conservation of the Universe by the term:
\begin{equation}
\dot{\rho} + 3 H (\rho + p) = 0.
\end{equation}
For the density and the pressure from equations (\ref{Friedmann1.2}) and (\ref{Friedmann2.2}) we get directly (\ref{scalar:Cosmo.2}), when the other matter fields are conserved as well:
\begin{equation}
\dot{\rho}_m + 3 H (\rho_m + p_m) = 0.
\end{equation}
So we end up with only one simple and consistent set of equations that depends on $\phi$ and $H$. Notice that we can simplify the equations using the redefinition
\begin{equation}\label{scalar:redefinition}
\tilde{V}(\phi) = V(\phi) + \frac{h_1(\phi)}{2f(\phi)}, \quad \tilde{h}(\phi) = \frac{h_2(\phi)}{2\sqrt{f(\phi)}},
\end{equation}
which simplifies the set into:
\begin{subequations}
\begin{equation}\label{scalar:Cosmo.2Sim}
\ddot{\phi} +3H\dot{\phi}\left(1 + \tilde{h}(\phi)\right) + \tilde{V}'(\phi) =0,
    \end{equation}
\begin{equation}\label{Friedmann1.2Sim}
3H^2=\frac{1}{2} \dot{\phi}^2 +\tilde{V}(\phi)
+\rho_m,
\end{equation}
\begin{equation}\label{Friedmann2.2Sim}
-3H^2-2\dot{H}=\frac{1}{2} \dot{\phi}^2 - \tilde{V}(\phi) + \dot{\phi}\tilde{h}(\phi) + p_m.
\end{equation}
\end{subequations}
If $h_2(\phi) = 0$, the set reduces to the canonical equations for a scalar field coupled minimally to gravity: the quintessence model \cite{Ratra:1987rm,Caldwell:1997ii,Zlatev:1998tr}. However, the non-trivial $\tilde{h}(\phi)$ term that is present only in the pressure equation, compensate energy density with the other components. Therefore, choosing the function yields the exact solution for the system.

\section{Dark energy, dark matter and stiff matter}\label{sec5}
In this section we test the impact of the second interaction in \eqref{L_int.di} with the function $h_2(\phi)$. The interaction couples the derivative of the scalar field $\phi$ into the vector field $u_{\mu}$. We consider constant functions:
\begin{equation}
\begin{split}
V(\phi) &= V, \quad f(\phi) = f, \\ h_1(\phi) &= h_1, \quad h_2(\phi) = h_2.\label{constfunc}
\end{split}
\end{equation}
The exact solution of Eq. (\ref{scalar:Cosmo.2}) for the scalar field is
\begin{equation}
\dot{\phi} = -\frac{C_0}{a^3}-\frac{h_2}{2 \sqrt{f}},\label{dotphi}
\end{equation}
where $C_0 > 0$ is a constant of integration. Inserting the relation \eqref{dotphi} into the Friedmann equations (\ref{Friedmann1.2}) and (\ref{Friedmann2.2}) yields:
\begin{subequations}
\begin{equation}
3 H^2 = \frac{C_0^2}{2a^6} + \frac{C_0 h_2}{2 \sqrt{f} a^3} + V + \frac{h_2^2 - 4 h_1}{8 f} + \rho_m,
\end{equation}
\begin{equation}
-3 H^2 - 2 \dot{H} = \frac{C_0^2}{2a^6} - V - \frac{h_2^2 - 4 h_1}{8 f} + p_m.
\end{equation}
\end{subequations}
There are three fluid components in addition to the
baryonic matter and radiation: dark matter ($w=0$),
\begin{equation}
 \rho_{DM} = \frac{C_0 h_2}{2 \sqrt{f} a^3}, \quad p_{DM} = 0,
\end{equation}
dark energy ($w = -1$),
\begin{equation}
 \rho_{DE} = V + \frac{h_2^2 - 4 h_1}{8 f}, \quad p_{DE} = -\rho_{DE} ,
\end{equation}
and stiff matter with equation of state parameter $w = 1$,
\begin{equation}
 \rho_{s} = p_{s} = \frac{C_0^2}{2a^6}.\label{rho_stiff}
\end{equation}
The stiff equation of state in cosmology is discussed in \cite{Zeldovich:1962emp,Chavanis:2014lra,Odintsov:2017cfr}. Since the density and pressure of the stiff matter component evolves as $a^{-6}$, which is a smaller power than the radiation solution $\sim a^{-4}$, we expect that this part would exist before radiation domination and decays rapidly in our universe. Moreover, some theories of the reheating mechanism require the kination domination era. Therefore, this picture of dark energy, dark matter and stiff matter will predict different scenarios with different potentials.

The coincidence problem rises the question why observable values of dark energy and dark matter densities in the late universe are of the same order of magnitude. The dark matter and the dark energy parts in this model are connected to each other through the parameters $f$ and $h_2$. This connection may explain the coincidence problem.

There is a correlation between the shift symmetry and an interaction between the components in the Universe. For constant functions the action has a hidden shift symmetry, that produces a Noether Symmetry:
\begin{equation}
\phi \rightarrow \phi +\mathrm{Const.} \quad\Rightarrow\quad j^{\mu} = \frac{1}{2} \phi^{,\mu} - \frac{1}{2}h_2(\phi) u^{\mu},
\end{equation}
which is covariantly conserved $\nabla_{\mu} j^{\mu} = 0$. When we introduce a nonconstant potential $V(\phi)$ or the function $\tilde{h}(\phi)$, the current is no longer conserved and an interaction between the components emerges. This correspondence was introduced in \cite{Guendelman:2012gg,Guendelman:2015rea} for a unified dark energy and dark matter models with modified measures. But the principle applies here and also for the Mimetic DM models.

\begin{figure*}[t!]
 	\centering
\includegraphics[width=1\textwidth]{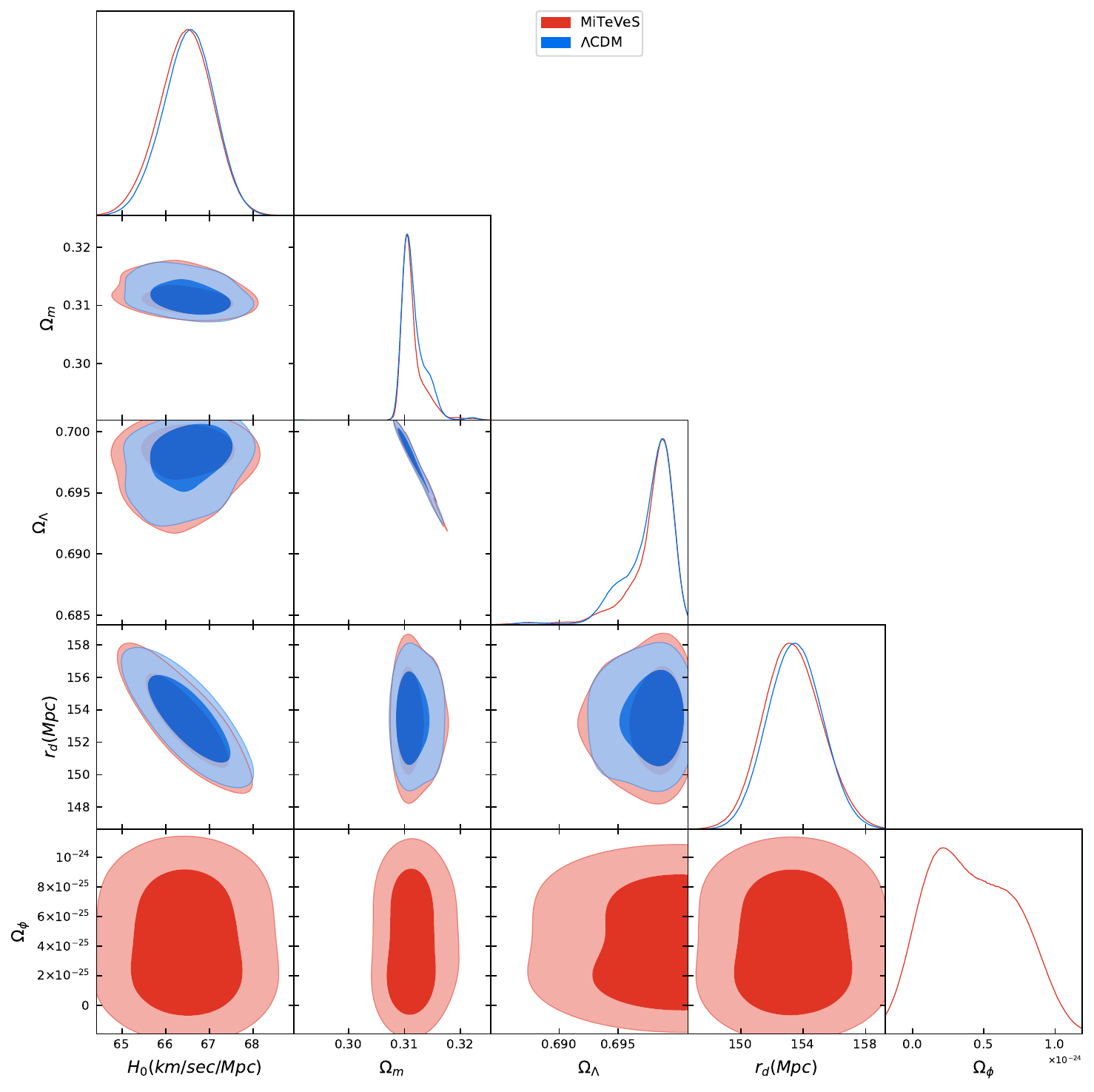}
%\begin{table*}
\begin{tabular}{| c | c | c | c | c | c | c | c |}
\hline\hline
Model & $H_0 (Km/sec/Mpc)$ & $\Omega_m$ & $\Omega_\Lambda$ & $\Omega_\phi (10^{-25})$   & $r_s (Mpc)$   & $ \Delta B_{ij}$\\
\hline\hline
MiTeVeS & $66.52 \pm 0.532$ & $0.31 \pm 0.021$ & $0.698 \pm 0.019$ & $4.258 \pm 2.87$ & $153.2 \pm 1.69$  &  $0.24$ \\
\hline
$\Lambda$CDM &  $66.58 \pm 0.47$ & $0.311 \pm 0.016$  & $0.698 \pm 0.01$ & - & $153.3 \pm 1.465$ & - \\
\hline\hline
\end{tabular}
\caption{\it{The posterior distribution for the simplest case of the MiTeVeS with constant functions (red curve) and for $\Lambda$CDM model (blue curve). The ratios of the matter density $\Omega_m$, dark energy $\Omega_\Lambda$, radiation density $\Omega_r$ and the stiff part $\Omega_\phi$ summarize to one. The final results for cosmological parameters for the MiTeVeS and the $\Lambda$CDM models are summarized in the table. In order to compare the models we calculate the Bayes factor.}}
%\end{table*}
\label{fig:fit}
\end{figure*}

\section{Cosmological Constraints}\label{sec6}
In order to constrain the additional equation of state we use standard candles (SCs), Baryon Acoustic Oscillations (BAO) and the Cosmic Microwave Background (CMB) distance priors (three data points that contain information of certain features of the power spectrum such as the position of a peak). Radiation is included in order to constraint the components also in the early universe. The normalized Hubble function reads:
\begin{equation}
\left(\frac{H(z)}{H_0}\right)^2 = \Omega_\Lambda + \Omega_m (1+z)^3 +  \Omega_r (1+z)^4 + \Omega_\phi (1+z)^6.
\label{eq:final}
\end{equation}
where $H_0$ is the current value of the Hubble parameter, and the other constants related to the MiTeVeS model are
\begin{equation}
\begin{split}
\Omega_\phi = \frac{C^2_0}{6 H_0^2}, \quad \Omega_m = \frac{C_0 h_2}{3\sqrt{f}H_0^2} +  \Omega_b, \\
\Omega_\Lambda = \frac{V}{3 H_0^2} + \frac{h_2^2 - 4 h_1}{24 f H_0^2},
\end{split}
\end{equation}
where $\Omega_\phi$ is the density parameter of the stiff matter, $\Omega_m$ is the density parameter of usual matter and dark matter, and $\Omega_\Lambda$ is the density parameter of dark energy. $\Omega_r$ is the density parameter of radiation. The matter density incorporates the dark matter density from the MiTeVeS fields as $C_0 h_2/3\sqrt{f}H_0^2$ and the density of baryons $\Omega_b$.
\begin{figure*}[t!]
\centering
\includegraphics[width=0.9\textwidth]{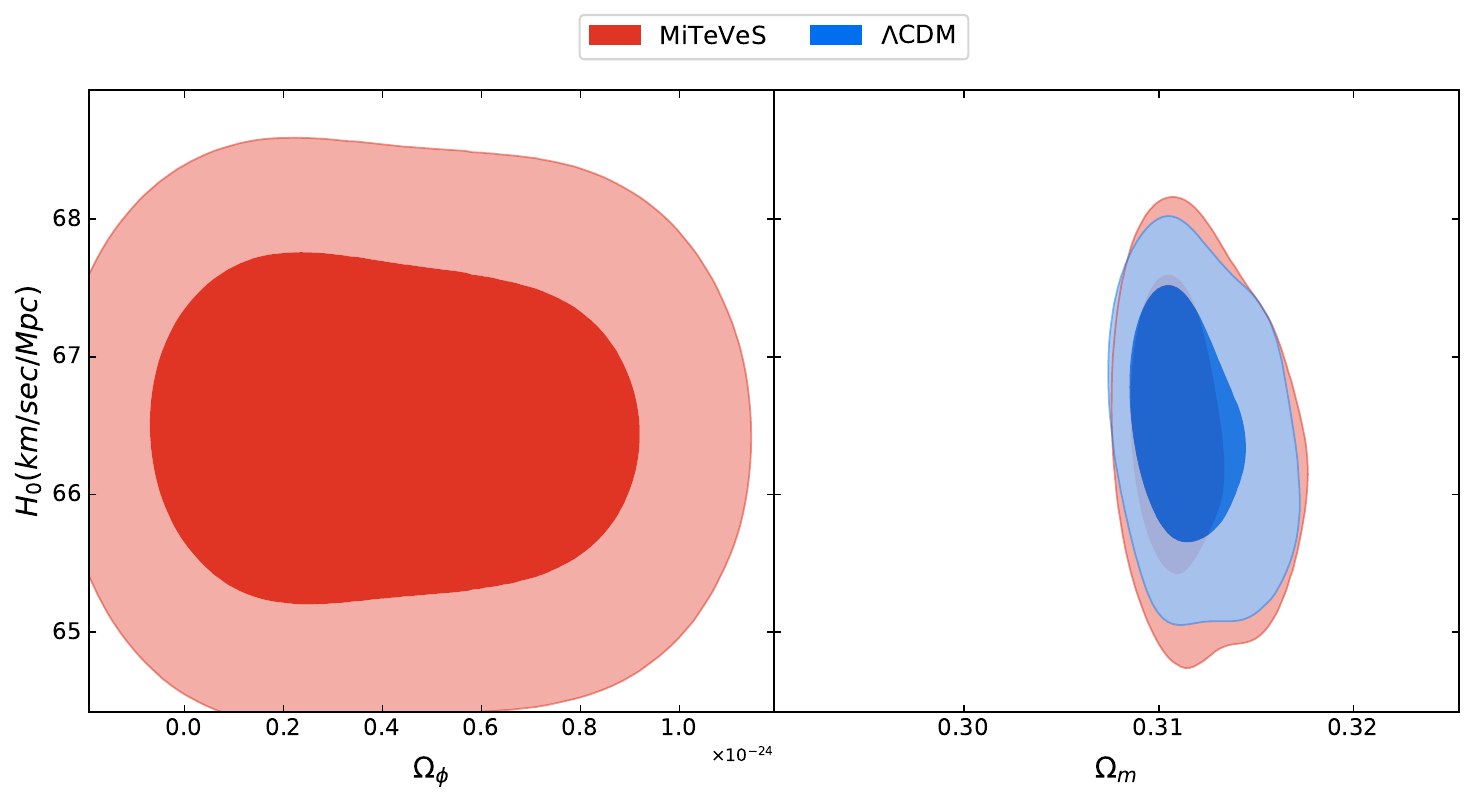}
\caption{\it{The posterior distribution for the simplest case of the MiTeVeS with constant functions (red curve) and for $\Lambda$CDM model (blue curve), with $H_0$ vs. $\Omega_m$.}}
\label{fig:fit2}
\end{figure*}

In order to constrain our model, we use a few data sets. First we impose the average bound on the possible variation of the Big Bang Nucleosynthesis (BBN) speed-up factor, defined as the ratio of the expansion rate predicted in a given model versus that of the $\Lambda$CDM model at the BBN epoch ($z_{BBN} \sim 10^9$). This amounts to the limit \cite{Uzan:2010pm} with
$\left({\Delta H}/{H_{\Lambda CDM}}\right)^2<0.01$, where $H_{\Lambda CDM}$ is the Hubble rate for the $\Lambda$CDM model and the $\Delta H$ is the difference between the MiTeVeS Hubble rate and the $\Lambda$CDM Hubble rate, $\Delta H=H-H_{\Lambda CDM}$. Imposing the relation (\ref{eq:final}) gives a bound on $\Omega_\phi$:
\begin{equation}
\Omega_\phi(Max) = \frac{\Omega_\Lambda + \Omega_m  (1+z_{BBN})^3 + \Omega_r  (1+z_{BBN})^4}{100 (1+z_{BBN})^6},\label{BBN-constraint}
\end{equation}
which gives a range of $10^{-22}$--$10^{-30}$ for the Planck values of $\Omega_m$ and $\Omega_\Lambda$. Therefore we set the value of $\Omega_\phi$ during BBN to be between zero and $\Omega_\phi(Max)$.

The Cosmic Chronometers (CC) exploit the evolution of differential ages of passive galaxies at different redshifts to directly constrain the Hubble parameter \cite{Jimenez:2001gg}. We use uncorrelated 30 CC measurements of $H(z)$ discussed in \cite{Moresco:2012by,Moresco:2012jh,Moresco:2015cya,Moresco:2016mzx}.

For Standard Candles (SC) we use measurements of the Pantheon Type Ia supernova dataset \cite{Scolnic:2017caz} from different binnes. The parameters of the model are fitted by comparing the observed value $\mu _{i}^{obs}$ to the theoretical value $\mu _{i}^{th}$ of the distance moduli, which is given by:
\begin{equation}
 \mu=m-M=5\log _{10}(D_{L})+\mu _{0},
\end{equation}
where $m$ and $M$ are the apparent and absolute magnitudes and $\mu
_{0}=5\log \left( H_{0}^{-1}/Mpc\right) +25$ is the nuisance parameter that
has been marginalized. The distance moduli is given for different redshifts $\mu_i=\mu(z_i)$. The luminosity distance is defined by
\begin{eqnarray}
D_L(z) &=&\frac{c}{H_{0}}(1+z)\int_{0}^{z}\frac{dz^{\ast }}{%
E(z^{\ast })}.
\end{eqnarray}%
Here, we assume that $\Omega _{k}=0$ (flat space-time). We use uncorrelated data points from different Baryon Acoustic Oscillations (BAO) \cite{Percival:2009xn,Beutler:2011hx,Busca:2012bu,Anderson:2012sa,Seo:2012xy,Ross:2014qpa,Tojeiro:2014eea,Bautista:2017wwp,deCarvalho:2017xye,Ata:2017dya,Abbott:2017wcz,Molavi:2019mlh}. Studies of the BAO feature in the transverse direction provide a measurement of $D_H(z)/r_d = c/H(z)r_d$, where $r_d$ is the sound horizon at the drag epoch and it is taken as an independent parameter and with the comoving angular diameter distance being:
\begin{equation}
D_M= \int_0^z\frac{c \, dz'}{H(z')}.
\end{equation}
In our database we also use the angular diameter distance $D_A=D_M/(1+z)$ and $D_V(z)/r_d$, which is a combination of the BAO peak coordinates above, namely
\begin{equation}
    D_V(z) \equiv [ z D_H(z) D_M^2(z) ]^{1/3}.
\end{equation}
The BAO data represent the absolute distance measurements in the Universe. From the measurements of correlation function or power spectrum of large scale structure, we can use the BAO signal to estimate the distance scales at different redshifts. The BAO data are analyzed based on a fiducial cosmology and the sound horizon at the drag epoch.

For the CMB data, we use the CMB shift parameters~\citep{Chen:2018dbv}. The values of the shift parameters with the corresponding covariant forms are called the CMB distant priors. The parameters read:
\begin{equation}
R = \sqrt{\Omega_m} H_0 r(z_{*})/c, \quad l_a = \pi r(z_*)/r_s(z_*),
\end{equation}
which are based on the Planck data release \citep{Aghanim:2018eyx}. where $r_{s}(z)$ is the comoving sound horizon and $z_{*}$ is the redshift with respect to the photon-decoupling surface. The comoving sound horizon is defined as
\begin{equation}
r_{s}(z) =  \frac{c}{H_{0}}\int_{0}^{a}\frac{da'}{\sqrt{3(1+\bar{R_{b}}a')a'^{4}E^{2}(z')}}.
\end{equation}
It is determined by the matter-radiation equality relation and $z_{\text{eq}}=2.5\times10^{4}\Omega_{m}h^{2}(T_{\text{CMB}}/2.7\text{K})^{-4}$ and with $\bar{R_{b}}=31500w_{b}(T_{\text{CMB}}/2.7\,\text{K})^{-4}$, where the CMB temperature is $T_{\text{CMB}}=2.7255\,\text{K}$.

Regarding the problem of likelihood maximization, we use an affine-invariant Markov Chain Monte Carlo (MCMC) sampler, as it is implemented within the open-source package Polychord \cite{Handley:2015fda} with the GetDist package \cite{Lewis:2019xzd} to present the results. The prior we choose is with a uniform distribution, where $\Omega_{m} \in [0.;1.]$, $\Omega_{\Lambda}\in[0.;1 - \Omega_{m}]$, $\Omega_{\phi}\in[0.;1 - \Omega_{m} - \Omega_{\Lambda}]$, $\Omega_{\phi}\in[0.;1 - \Omega_{\phi } (Max)]$, $\Omega_{r}\in[0.;1 - \Omega_{m} - \Omega_{\Lambda} - \Omega_{\phi}]$, $H_0\in [50;100]$ km/sec/Mpc. For the baryonic matter we use the value reported by Planck 2018.

The posterior distribution is presented in Fig.~\ref{fig:fit}, which describes the best fit of the MCMC analysis. The differences between the parameters of MiTeVeS and $\Lambda$CDM are rather small. For the MiTeVeS the Hubble parameter is $66.52 \pm 0.53$ km/sec/Mpc and for the $\Lambda$CDM model $66.58 \pm 0.47$ km/sec/Mpc. The matter density is $0.31 \pm 0.021  $ for MiTeVeS and $0.311 \pm 0.016$ for $\Lambda$CDM. The dark energy density is $0.698 \pm 0.019 $ for MiTeVeS and $0.698 \pm 0.01 $ for the $\Lambda$CDM model. Because of the error bars of all of these parameters, one cannot distinguish the difference from the current cosmological measurements. However, according to current measurements, the stiff matter may exist in our Universe, but its density portion is constrained to be of order $10^{-25}$ or smaller. To be more precise, the MCMC gives:
\begin{equation}
\Omega_\phi = \left(4.258 \pm 2.877\right) \cdot 10^{-25}.
\label{eq:upperbound}
\end{equation}
The value is compatible with zero with $2\sigma$. In order to test the preference of the models, we use the Bayesian Evidence that is calculated from the Polychord directly \cite{doi:10.1080/01621459.1995.10476572}. The difference between the Bayesian Evidence yields $0.24$, which is an indistinguishable preference for the $\Lambda$CDM model.

Although the value of the stiff matter density $\Omega_\phi$ has to be very small in order to be consistent with the cosmological observations, being constrained below a maximum value during BBN \eqref{BBN-constraint} and having the present value \eqref{eq:upperbound} in our model, stiff matter may play a role in the very early universe. Also it could be interesting to mention that a cosmological fluid with the stiff equation of state appears in gravitational theories where the connection has torsion \cite{Lucat:2015rla,Unger:2018oqo,Medina:2018rnl}.
Torsion appears in several types of gravitational theories, such as in the Einstein-Cartan and the gauge theories of gravity (for some comprehensive reviews, see, e.g. \cite{Hehl:1994ue,Blagojevic:2013xpa} and the references therein) as well as in supergravity (see for example \cite{VanNieuwenhuizen:1981ae}).\footnote{We thank Friedrich Hehl for several discussions regarding the torsion.}
If the stiff fluid has a negative energy density, which is the the case for example in the Einstein-Cartan theory of gravity \cite{Medina:2018rnl}, there is no initial singularity \cite{Chavanis:2014lra,Trautman:1973wy}, since the scale factor remains nonzero at the beginning. In our model, when there are no interactions between stiff matter, dark matter and dark energy due to the choice of constant functions \eqref{constfunc}, the energy density of stiff matter is positive \eqref{rho_stiff}, and hence the initial singularity exists \cite{Chavanis:2014lra} similarly as in the $\Lambda$CDM model.

\begin{figure}[t!]
\centering
\includegraphics[width=0.45\textwidth]{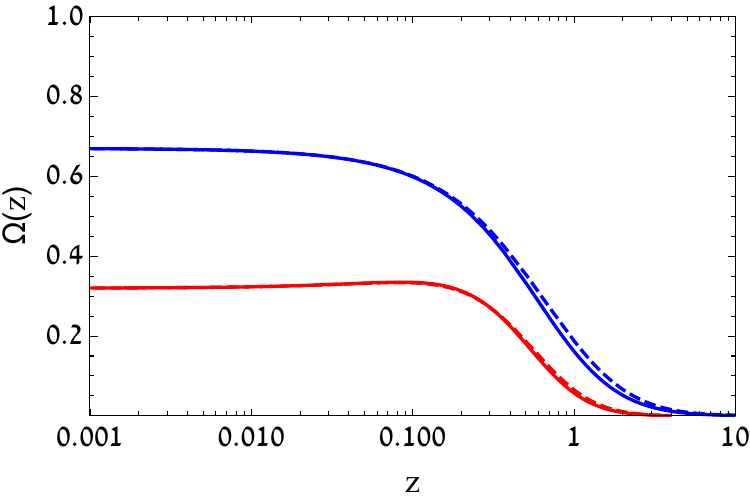}
\caption{\it{The evolution of the $\Omega_m$ (red) and $V(\phi) = \Omega_\Lambda$ (blue) for low and high redshifts, without any interaction (smooth line) or with interaction where $\beta = 1$ (smooth line).  }}
\label{fig:eval}
\end{figure}

\section{Interacting dark energy and dark matter}\label{sec7}
In this section we discuss the further possibilities of the MiTeVeS theory with general interaction forms. For the solution discussed in sections~\ref{sec5} and \ref{sec6} the mimetic dark matter component is not distinguishable from dust at the background level. However, for nonconstant functions (Eq.~\ref{constfunc}) one can get interacting dynamical dark energy and dark matter, which emerges naturally from the MiTeVeS action.

In order to track the evolution of the Universe (numerically and analytically) in the general case we rewrite the field equations in terms of the redshift instead of time, which is done using the relation:
\begin{equation}
\frac{d}{dt} = - H (z) \left(1+z\right) \frac{d}{dz}.
\end{equation}
The modified scalar field equation is given as:
\begin{equation}
\begin{split}
\phi'(z) \left((z+1) \frac{H'(z)}{H}- (3\tilde{h}(\phi)+2)\right) \\ +(z+1) \phi''(z)  +\frac{V'(\phi)}{(z+1) H^2}=0,
\end{split}
\end{equation}
where $H = H(z)$ and $\phi = \phi (z)$ are functions of the redshift. In principle, we need two equations to describe the full evolution. Equations (\ref{Friedmann1.2Sim}) and (\ref{Friedmann2.2Sim}) describe the evolution of $\phi(z)$ and $E(z) = H(z)/H_0$:
\begin{subequations}
\begin{equation}
\begin{split}
E'(z) &= \frac{3}{2} \tilde{h}(\phi) \phi'(z) \\
&+\frac{3 E(z)^2-3 \tilde{V}(\phi) - 1.5 \Omega_m(z)+1.5 \Omega_p(z)}{E(z) (1 + z)},
\end{split}
\end{equation}
\begin{equation}
\phi'(z)^2 = \frac{E(z)^2-\tilde{V}(\phi)-\Omega_m(z)}{(1 + z)^2 E(z)^2},
\end{equation}
\end{subequations}
where $\Omega_m(z) = \rho_m/3H_0^2$ and $\Omega_p(z) = p_m/3H_0^2$. In order to determine the evolution of the variables for a general form of the functions, $E(z)$ and $\phi(z)$ are evaluated via these two equations. For the case $\tilde{h}(z) = 0$ the system recovers to the standard solution, where $\dot{\phi}^2 \sim (1+z)^{-3}$.

A nonconstant function $\tilde{h}(\phi)$ can give an interaction between dark energy and dark matter. As an example we take: $\tilde{h}(\phi) = \beta \phi$. The numerical evaluation of the matter density (in red) and the dark energy density (in blue) is described in Fig.~\ref{fig:eval}. The smooth line shows the evaluation without any interaction and the dashed line shows the evaluation with interaction, where we assume $\beta = 1$ and $\phi(z=0) = 0$ for simplicity. The interaction changes the rates between the dark energy and dark matter. The possibility for such an interaction appears naturally in the MiTeVeS action.

\section{Discussion and conclusion}

In this paper we have uncovered some cosmological implications of the MiTeVeS theory. The particular formulation of the theory includes both a scalar field and a vector field, which is stable and free from ghosts. The two simplest interactions between the scalar field and the vector field were considered.
In the homogeneous FLRW background, the first interaction in \eqref{L_int.di} results only to a redefinition of the potential of the scalar field \eqref{scalar:redefinition}. However, the second interaction in \eqref{L_int.di} has interesting cosmological implications.
For constant functions \eqref{constfunc}, a conserved Noether current emerges. From the scalar field dynamics, we obtained dark energy, dark matter and stiff matter. When the shift symmetry is broken, the current is not covariantly conserved. Consequently, possible interactions between these three components emerge.

We use the latest observations of the Big Bang Nucleosynthesis, Cosmic Chronometers, Type Ia Supernova, Baryon Acoustic Oscillations and the Cosmic Microwave Background distant prior to constrain the additional stiff matter contribution, which is not present in the standard $\Lambda$CDM model. It is clear that the dynamics of the Universe does not change significantly as long as the energy density of the stiff matter part is around $10^{-25}$ or smaller.

There are a few directions to continue the research. It is possible to explore different potentials and functions that approach constant functions asymptotically. We show one simple example of an interaction that changes the rates between dark energy and dark matter naturally from the action, which could be investigated in the future. Such functions produce interaction between the stiff matter, the dark matter and the dark energy parts, and may resolve the cosmic tensions regarding $H_0$ and others. Moreover, it is important to track the perturbation solutions for this theory in order to compare with further data sets and to validate the formulation in different systems.

The third interesting option is to explore the effect of the scalar $\Phi$ in this theory. This scalar is suggested in the theory and gives the behavior of dynamical $G \sim \Phi^{-2}$. Many versions of MOND and other modified theories of gravity suggest this effect that naturally emerges in this formulation. So the full theory gives additional effect of emerging gravity that should be studied in the future.

% \section*{Acknowledgments}
\begin{acknowledgments}
We would like to thank Eduardo I. Guendelman for helping us to conclude the solution with the constant functions case. D.B. thanks the Rothschild and the Blavatnik Fellowships for financial support. This work was done in a Short Term Scientific Missions (STSM) in Finland funded by the COST Action CA15117 ``Cosmology and Astrophysics Network for Theoretical Advances and Training Action''. M.O. gratefully acknowledges funding from the Oskar Öflunds Stiftelse.
\end{acknowledgments}

\bibliographystyle{apsrev4-1}
\bibliography{ref}

\end{document}